\documentclass[twocolumn,showpacs,preprintnumbers,amsmath,amssymb,floatfix]{revtex4}
\newcommand{\newwidth}{0.3375\textwidth}
\newcommand{\newheight}{0.225\textwidth}

\newcommand{\newwidthprime}{0.13\textwidth}
\newcommand{\newheightprime}{0.15\textwidth}
\usepackage{graphicx}
\usepackage{dcolumn,psfrag}
\usepackage{bm,verbatim}
\usepackage{color}

\begin{document}

\title{Cooling by heating in nonequilibrium nanosystems} 

\author{R.\ H\"artle$^{1,5}$}
\author{C.\ Schinabeck$^{2,5}$}
\author{M.\ Kulkarni$^{3}$}
\author{D.\ Gelbwaser-Klimovsky$^{4}$}
\author{M.\ Thoss$^{2,5}$}
\author{U.\ Peskin$^{6}$}
\affiliation{
$^1$ Institute of Theoretical Physics, University of Gottingen, 37077 G\"ottingen, Germany, \\
$^2$ Institute of Theoretical Physics and Interdisciplinary Center for Molecular Materials, 
University of Erlangen-Nuremberg, 91058 Erlangen, Germany,\\
$^3$ International Centre for Theoretical Sciences, Tata Institute of Fundamental Research, Bangalore - 560089, India,\\
$^4$ Department of Chemistry and Chemical Biology, Harvard University, Cambridge, MA 02138, USA,\\
$^5$ Institute of Physics, University of Freiburg, 79104 Freiburg,  Germany, \\
$^6$ Schulich Faculty of Chemistry, Technion-Israel Institute of Technology, Haifa 32000, Israel.
}

\date{\today}

\begin{abstract}
We demonstrate the possiblity to cool nanoelectronic systems in nonequilibrium situations 
by increasing the temperature of the environment. Such cooling by heating is possible for a variety of experimental conditions where the relevant 
transport-induced excitation processes become quenched and deexcitation processes 
are enhanced upon an increase of temperature. 
The phenomenon turns out to be robust with respect to all relevant parameters. It is especially pronounced for 
higher bias voltages and weak to moderate coupling. 
Our findings have implications for open quantum systems in general, where electron transport is coupled to mechanical 
(phononic) or photonic degrees of freedom. In particular, molecular junctions with rigid 
tunneling pathways or quantum dot circuit QED systems meet the required conditions.
\end{abstract}


\maketitle

Nanoelectronic systems exhibit a plethora of fundamentally interesting physical properties 
and, at the same time, are considered as promising architectures for technological applications, 
ranging from transistor \cite{Piva2005,Song2009} to quantum information devices \cite{Loss1998,Petersson2011}. 
Experimental realizations include single-molecule junctions 
\cite{Reed97,Reichert02,Park02,Reed2004,WuNazinHo04,Cuniberti05,Venkataraman06,Secker2010,cuevasscheer2010}, atomic wires 
\cite{Agrait2002,Agrait2003,Matzdorf2009,Matzdorf2011}, carbon nanotube 
\cite{LeRoy,Tans97,Sapmaz05,Sapmaz06,LeturcqOppen09,Delbecq2011,Delbecq2013} and 
semiconductor based quantum dot systems 
\cite{Reed1988,Wees1988,Holleitner2001,Wiel2002,Wegscheider2007,Kiesslich2007,Tutuc2011,Liu2014,Beckel2014}. 
A limiting factor is current-induced heating associated with the excitation of mechanical or 
electromagnetic degrees of freedom. 
Such heating limits the control, coherence or even the mechanical stability of these devices. 
It is therefore expedient to identify and understand the intrinsic cooling mechanisms of nanoelectronic systems.

\begin{figure}
\begin{center}
\begin{tabular}{ccc}
(a) \text{\footnotesize transport} &(b) \text{\footnotesize transport} &(c) \text{\footnotesize pair-creation}\\
\resizebox{\newwidthprime}{\newheightprime}{
\includegraphics{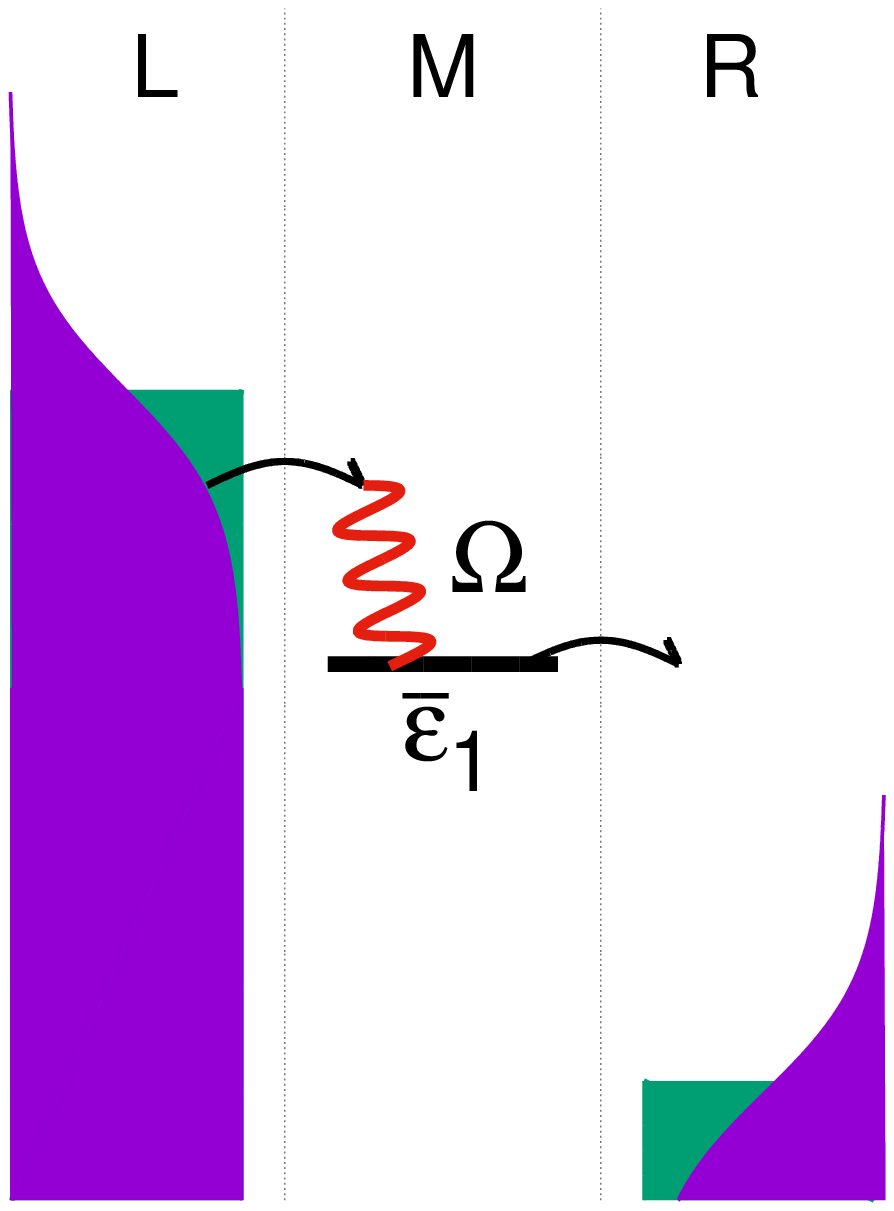}
}&
\resizebox{\newwidthprime}{\newheightprime}{
\includegraphics{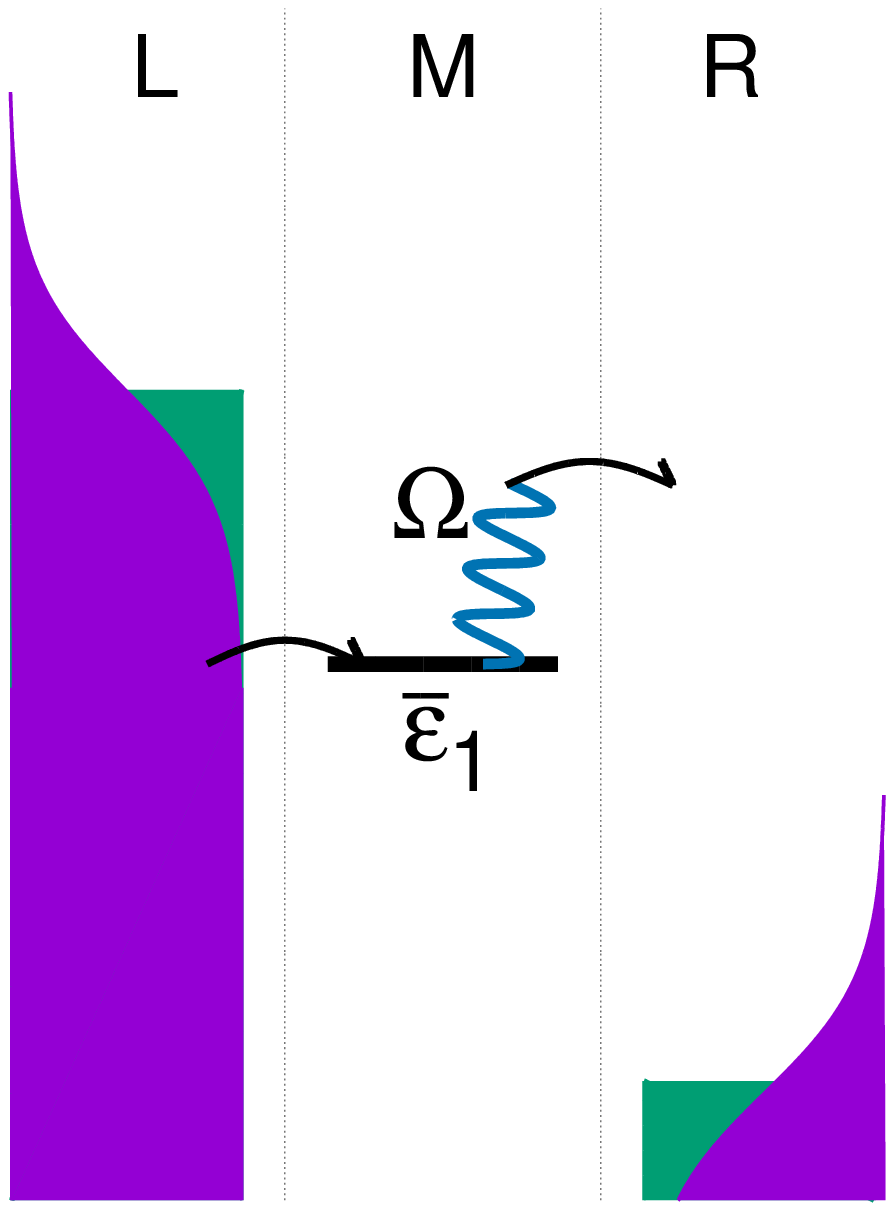}
}&
\resizebox{\newwidthprime}{\newheightprime}{
\includegraphics{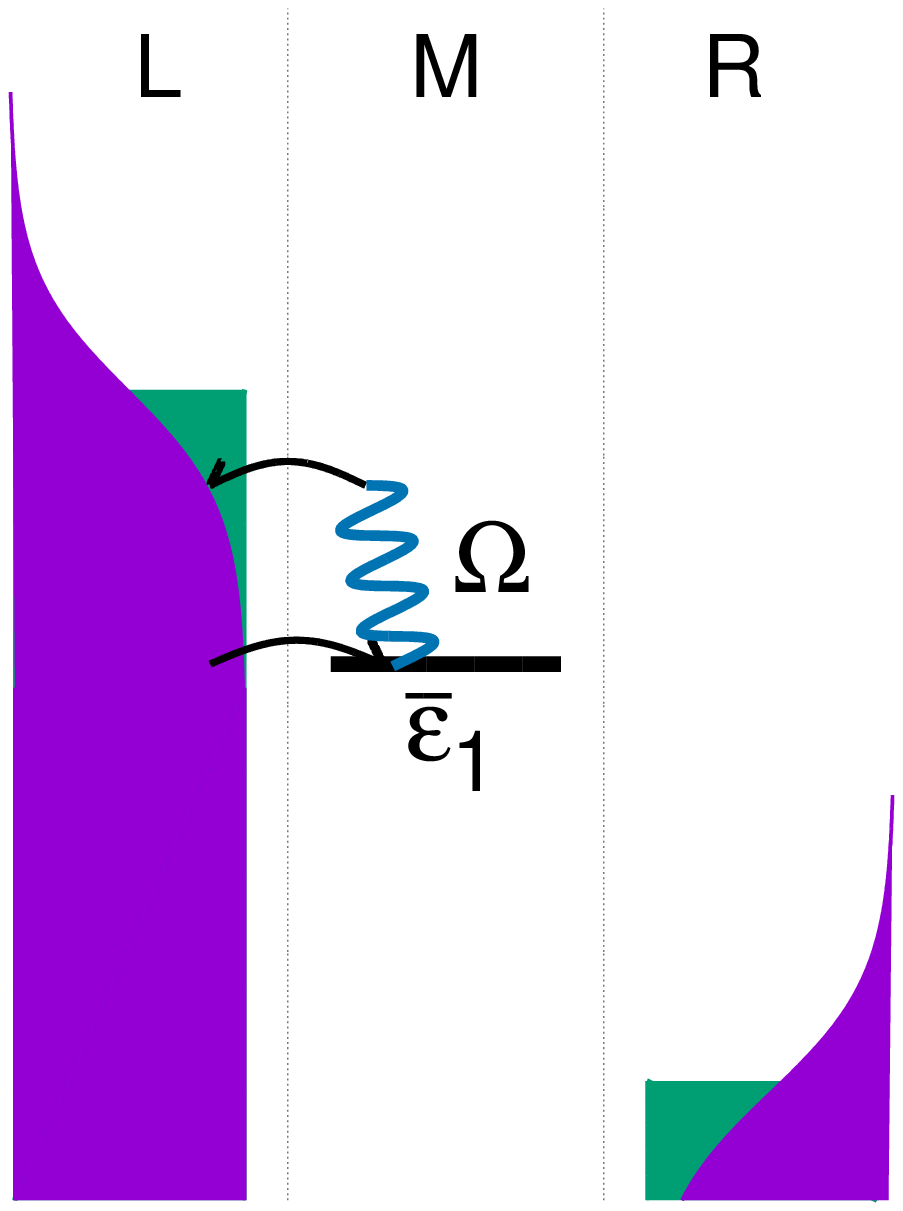}
}\\ 
\end{tabular}
\end{center}
\caption{\label{processes} Sketch of representative inelastic processes in a biased molecular junction, 
where electrons tunnel in two, sequential resonant tunneling processes onto and off the molecule (M). 
The green and purple areas represent occupied states in the left (L) and right lead (R) for lower and higher temperatures, respectively. 
Panel (a) and (b) depict transport processes, where 
the vibrational mode is excited and deexcited upon tunneling of an electron from L to M 
and from M to R, respectively. At low temperatures, process (a) requires higher bias voltages, 
$\Phi\gtrsim2(\overline{\epsilon}_1+\Omega)$ than process (b), $\Phi\gtrsim2\overline{\epsilon}_1$. At these voltages, 
an increase of temperature reduces the initial state population of both processes. 
Panel (c) depicts a pair-creation process, where the electron returns to the original electrode. 
For $\Phi\gtrsim2(\overline{\epsilon}_1+\Omega)$ and low temperatures, this deexcitation process 
is Pauli-blocked since its final state population is almost $1$. It can be reactivated 
by an increase of temperature which reduces the final state population. }
\end{figure}

Nonequilibrium steady states 
offer the possibility for unprecedented control or cooling strategies. One possibility is to use the bias 
voltage which is applied to the device. 
In many situations, a non-zero voltage leads to additional heating processes and higher levels of excitation. 
However, in the presence of co-tunneling assisted sequential tunneling \cite{Lueffe,Huettel2009}, 
higher-lying electronic states \cite{Hartle09,Romano10}, antiresonances \cite{McEniry2009}, suitably defined spectral 
properties of the leads \cite{Saito2009,Gelbwaser2017}, donor-acceptor structures \cite{Saito2009,Brandbyge2011,Simine2012} 
or electron-electron interactions \cite{Hartle2010b}, an increase in voltage can lead to lower excitation levels of the 
mechanical or photonic degrees of freedom. Another possibility is to use spin-polarized currents \cite{Bruggemann2014}.

In this work, we demonstrate how an increase of the environments temperature 
leads to lower excitation levels of the nanoelectronic system and thus stabilizes the device. 
To be specific, we consider models of single-molecule junctions in the following. The principle, however, applies to 
any few-level systems where electron transport is coupled to mechanical or, e.g.\ in the case of cavity QED 
systems \cite{Delbecq2011,Delbecq2013,Liu2014}, photonic degrees of freedom. 
It is emphasized that our scheme for cooling by heating does not involve heat or temperature gradients such as 
in domestic \cite{Gordon2000} or quantum absorption refrigerators 
\cite{Palao2001,Linden2010,Levy2012,Cleuren2012,Gelbwaser2014,Gelbwaser2015} or 
optomechanical devices \cite{Mari2012}. Rather, chemical potential differences induced by an external 
bias voltage lead to transport through the nanosystem between two Fermion reservoirs where heating the reservoirs 
cools down the nanosystem.

The phenomenon of thermal stabilization predicted here is most strongly pronounced in the regime of weak to 
moderate molecule-lead and electronic-vibrational/photonic coupling 
and high enough (but still practically relevant) voltages, where transport-induced 
heating mechanisms (cf.\ Fig.\ \ref{processes}a) are quenched for higher temperatures 
and, at the same time, cooling mechanisms are either less affected (cf.\ Fig.\ \ref{processes}b) 
or become enhanced (cf.\ Fig.\ \ref{processes}c). 
An important characteristics of this regime is that weaker coupling to vibrational or photonic degrees of 
freedom results in higher excitation levels 
\cite{Mitra04,Semmelhack,Avriller2010,Hartle2010b,Hartle2011,Ankerhold2011,Hartle2015b}. 
In the limit of vanishing couplings and temperatures, this leads to an indefinite increase of the excitation level 
for harmonic modes \cite{Hartle2011,Hartle2015b}, that is a vibrational instability \footnote{Note that the term vibrational 
instability is also used in a slightly different context where internal heating processes emerge from a donor-acceptor 
structure of the molecule \cite{Brandbyge2011,Simine2012}. }.

We consider transport through a molecule (M) that is coupled to a left (L) 
and a right electrode (R). We describe this transport setup by the Hamiltonian 
$H=H_{\text{M}}+H_{\text{ME}}+H_{\text{E}}$ where (using units where $\hbar=1$ and $k_\text{B}=1$)
\begin{eqnarray}
H_{\text{M}} &=& \sum_{m\in \text{M}}\epsilon_{m} d_m^{\dagger}d_m + \sum_{m<n\in\text{M}} U_{mn} d_m^{\dagger}d_m d_n^{\dagger}d_n \\
&& + H_{\text{Vib}} + \sum_{m\in \text{M}} \lambda_m (a+a^\dagger) d_m^{\dagger}d_m, \nonumber\\
H_{\text{Vib}} &=& \Omega a^\dagger a + \sum_{\xi} V_\xi (a+a^\dagger)^\xi, \\
H_{\text{E}} &=& \sum_{k\in\text{L,R}} \epsilon_{k} c_{k}^{\dagger}c_{k}+ \sum_{\alpha} \omega_{\alpha} b_{\alpha}^{\dagger}b_{\alpha}, \\
H_{\text{ME}} &=& \sum_{m\in \text{M},k\in \text{L,R}} ( V_{mk} c^{\dagger}_{k} d_m + V_{mk}^* d_m^{\dagger} c_{k}  ) \\
&& +\sum_{\alpha} W_{\alpha}  (a+a^\dagger) (b_\alpha+b_\alpha^\dagger). \nonumber
\end{eqnarray}
It includes a discrete set of electronic eigenstates with energies $\epsilon_{m}$ and 
density-density interactions $U_{mn}$ between electrons in states $m$ and $n$. 
The states of the molecule are coupled to a continuum of electronic states with energies $\epsilon_{k}$ 
and coupling matrix elements $V_{mk}$ in the left (L) and the right (R) electrode. 
The corresponding tunneling efficiency or hybridization function is 
$\Gamma_{K,mn}(\epsilon) = 2\pi \sum_{k\in K} V_{mk}^{*} V_{nk} \delta(\epsilon-\epsilon_{k})$ ($K\in\{\text{L,R}\}$). 
Throughout this work, we use the wide-band approximation $\Gamma_{K,mn}(\epsilon)\approx \Gamma_{K,mn}$ 
and consider a symmetric drop of the applied bias voltage, i.e.\ the chemical potentials in the left 
and right lead are given by $\mu_{\text{L/R}}=\pm \Phi/2$. 
We model vibrational effects by the vibrational Hamiltonian $H_{\text{Vib}}$,  
which, for $V_\xi=0$, is harmonic with frequency $\Omega$ and, for non-zero $V_{\xi}$, includes generic anharmonic effects, 
most importantly a non-equidistant energy spectrum.  
The mode may be representative of the dominant reaction coordinate of the molecule. 
It is coupled to the electronic states of the molecule by coupling strengths $\lambda_{m}$ and 
a bath of harmonic oscillators $\alpha$ with coupling strengths $W_{\alpha}$ 
in order to account for intramolecular 
vibrational energy redistribution \cite{Mukamel1980,Freed,Zewail,Nesbitt1996} 
or energy losses to a bosonic junction's environment \cite{Tao2006,Seidemann10}. 
Note that we use the same temperature $T$ to describe the bath and the electrodes, 
that is we do not consider any external temperature gradients.

We solve the above transport problem using the well established Born-Markov (BM)
\cite{May02,Mitra04,Lehmann04,Harbola2006,Volkovich2008,Hartle09,Hartle2010b,Pshenichnyuk2010} 
and the recently developed hierarchical quantum master equation (HQME) approach \cite{Schinabeck2016}. 
Explicit formulas and detailed derivations for HQME and BM can be found in Refs.\ \cite{Jin2008,Schinabeck2016} 
and \cite{Hartle2010b,Hartle2011}, respectively. The central quantity of both approaches 
is the reduced density matrix $\sigma$ of the molecule which includes the electronic levels and the vibrational mode. 
We determine it as the stationary solution of its equation of motion,  
\begin{eqnarray}
0\,\stackrel{!}{=}\,\frac{\partial \sigma(t)}{\partial t} \,=\, 
-i \left[ H_{\text{M}} , \sigma(t) \right] - \sigma^{(1)}(t), 
\end{eqnarray}
where the first term of the rhs describes the internal dynamics of the molecule 
and the second term $\sigma^{(1)}(t)$ the influence of the environment. 
Using HQME, we determine $\sigma^{(1)}(t)$ by solving its equation of motion, which leads 
to second- and, consequently, higher-tier operators $\sigma^{(n)}(t)$ with $n>1$. 
Truncation of this hierarchy corresponds to a truncated hybridization expansion. 
HQME thus allows us to systematically assess the importance of higher-order effects. The basic effect, however, is 
already included in BM, where the first-tier operator 
\begin{eqnarray}
\sigma^{(1)}(t) &=& \int_{0}^{\infty} \text{d}\tau\, 
\text{tr}_{\text{E}}\lbrace \left[ H_{\text{ME}} , \left[ H_{\text{ME}}(\tau), \sigma(t) 
\sigma_{\text{E}} \right] \right] \rbrace 
\end{eqnarray} 
is expressed in terms of the reduced density matrix $\sigma(t)$, the density matrix of the 
environment $\sigma_{\text{E}}$ and the coupling operator $H_{\text{ME}}(\tau)= 
\text{e}^{-i(H_{\text{M}}+H_{\text{E}})\tau} H_{\text{ME}} 
\text{e}^{i(H_{\text{M}}+H_{\text{E}})\tau}$. 
Thus, BM corresponds to a first-tier truncation of HQME, including 
the Markov approximation $\sigma(t-\tau)\approx \text{e}^{iH_{\text{M}}\tau} \sigma(t) 
\text{e}^{-iH_{\text{M}}\tau}$ which enters the above integral kernel. 
In addition, we disregard off-diagonal elements of the density 
matrix and renormalization effects. A discussion on the role of vibrational off-diagonal elements 
can be found, for example, in Ref.\ \cite{Hartle2010b}. The role of renormalization effects 
in the presence of electron-electron interactions has been outlined in 
Refs.\ \cite{Martinek2003,Wunsch2005,Hartle2014}, including a pronounced resonance 
in the conductance-voltage characteristics \cite{Hartle2013b,Hell2014,Wenderoth2016}. These effects, 
however, are not important for our discussion, which we show explicitly by a comparison to HQME 
where both off-diagonal elements and renormalization effects are included.

We introduce the phenomenon of thermal stabilization for a minimal model first. 
It includes a single electronic state with the 
polaron-shifted energy $\overline{\epsilon}_1\equiv\epsilon_1-\lambda_1^2/\Omega=3\Omega$  
that is weakly coupled to an undamped ($W_{\alpha}=0$) harmonic mode 
with $\lambda_{1}=\Omega/10$ and the leads with $\Gamma_{\text{L/R},11}=\Omega/1000\equiv\Gamma_{\text{L/R}}$. 
We measure the stability of the junction by the average vibrational energy $\langle H_{\text{Vib}} \rangle$. 
It is the result of thermal excitations, 
transport-induced excitation processes (Figs.\ \ref{processes}a) and 
deexcitation processes which are associated either with transport (Fig.\ \ref{processes}b) or 
pair-creation processes (Fig.\ \ref{processes}c). High values indicate a less stable junction, 
low ones a more stable junction.

\begin{figure}
\resizebox{0.5\textwidth}{0.45\textwidth}{
\hspace{-2.cm}\includegraphics[scale=1]{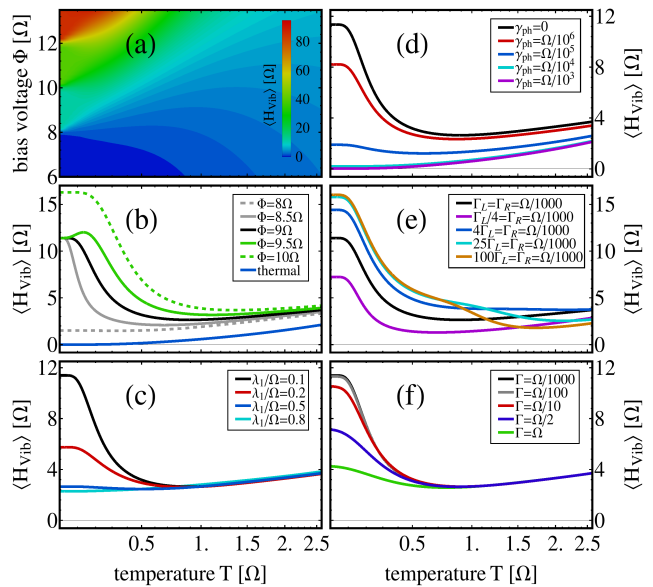}
}
\caption{\label{basics} Average vibrational energy $\langle H_{\text{Vib}} \rangle$ for our model of a molecular junction 
as a function of both temperature and applied bias voltage $\Phi$ (Panel (a)), as a function of temperature 
for fixed voltages (Panel (b)), different 
electronic-vibrational coupling strengths $\lambda$ (Panel (c)), 
mode-bath couplings (Panel (d)), molecule-lead coupling 
ratios $\Gamma_{\text{L}}/\Gamma_{\text{R}}$ (Panel (e)) 
and strengths $\Gamma$ (Panel (f)). 
For the results shown in Panels (c)--(f), we used $\Phi=9\Omega$. If not stated otherwise, 
the other parameters are $\overline{\epsilon}_1=3\Omega$, $\lambda_1=\Omega/10$, 
$\Gamma_{\text{L}}=\Gamma_{\text{R}}=\Omega/1000$ and $\gamma_{\text{ph}} = 
2\pi \sum_{\alpha} \vert W_{\alpha} \vert^{2} \delta(\Omega-\omega_{\alpha})$. 
The results shown in Panels (a)--(e) and (f) have been obtained using BM and HQME, respectively. 
}
\end{figure}

The average vibrational energy $\langle H_{\text{Vib}} \rangle$ of our minimal model is shown in Fig.\ \ref{basics}a as a function of both 
bias voltage $\Phi$ and temperature $T$. At very low temperatures, $T\rightarrow0$, 
we observe a step-like increase with bias voltages. The steps at $\Phi=2(\overline{\epsilon}_1+\Omega)=8\Omega$, 
$10\Omega$, $12\Omega$ ... correspond to the opening of transport-induced heating processes (see Fig.\ \ref{processes}a) 
and the closing of cooling processes via pair creation (see Fig.\ \ref{processes}c), 
which are associated with single, two, three ... phonon transitions, respectively. 
Increasing the temperature, the steps are smeared out and the vibrational energy decreases due to suppression (enhancement) 
of heating (cooling) rates as explained below. 

We study this temperature effect more closely in Fig.\ \ref{basics}b 
which shows the vibrational energy as a function of temperature for different fixed 
values of the applied bias voltage $\Phi$. 
For low voltages $\Phi\lesssim2(\overline{\epsilon}_1+\Omega)$ (dashed grey line), 
the vibrational energy increases monotonically with temperature, starting from low transport-induced values which evolve 
towards values that are given by a thermal distribution (cf.\ the blue line which depicts 
the Bose function $1/(\text{exp}(\Omega/T)-1))$). 
The vibrational energy is relatively low in this regime because transport-induced heating 
is outbalanced by both transport-induced cooling (Fig.\ \ref{processes}b) and pair-creation processes 
(Fig.\ \ref{processes}c) \cite{Hartle2010b}. 
Additional heating processes (Fig.\ \ref{processes}a) become active 
at higher bias voltages $\Phi\gtrsim2(\overline{\epsilon}_1+\Omega)$, 
while, in parallel, pair-creation processes with a single phonon transition become blocked (e.g.\ the one depicted in Fig.\ \ref{processes}c). 
Thus, the transport-induced vibrational energy increases significantly at these voltages (cf.\ solid and dashed green lines). This leads to a 
qualitatively different temperature dependence, in particular a negative slope in an intermediate temperature regime, 
which becomes more pronounced for higher bias voltages. This is the regime where thermal stabilization occurs, 
that is a decrease of the vibrational energy as the temperature of the environment is increased. The mechanism can be 
rationalized by the effect of thermal broadening which quenches preferentially 
the additional heating processes (compare Fig.\ \ref{processes}a to Fig.\ \ref{processes}b) and 
allows pair-creation processes (Fig.\ \ref{processes}c) to cool the junction again 
before thermal fluctuations override any transport-induced effects. 
For very low temperatures, $T\ll\Omega/2$, these broadening effects are too inefficient, 
as reflected by the plateau which appears before the average vibrational energy decreases. 
The small peak in the green line indicates the onset of transport-induced 
excitation processes (Fig.\ \ref{processes}a) and the suppression of pair creation processes (Fig.\ \ref{processes}c) 
with two vibrational quanta which, for $\Phi=9.5\Omega$, 
occurs before pair-creation processes with a single vibrational quantum 
become re-activated.

As will be shown below, the phenomenon of thermal stabilization is robust. Nevertheless, the extent of 
the regime where thermal stabilization occurs depends on the specific junction parameters 
which we now study one by one. 
To this end, we point out that the phenomenon occurs in the same regime of bias voltages, 
$\Phi\gtrsim2(\overline{\epsilon}_1+\Omega)$, 
where the model exhibits a vibrational instability as $\lambda,T,\Gamma_{\text{L/R}}\rightarrow0$, 
that is an indefinite increase of vibrational energy as the electronic-vibrational coupling constant 
is decreased \cite{Mitra04,Semmelhack,Avriller2010,Ankerhold2011,Hartle2011,Hartle2015b}. 
Consequently, the transport-induced vibrational energy is higher and the stabilization effect is more pronounced 
for weaker electronic-vibrational coupling, as can be seen in Fig.\ \ref{basics}c, 
where we show the temperature dependence of the vibrational 
energy for different electronic-vibrational coupling strengths. 
While the vibrational energy of the mode is indeed lower for stronger coupling strengths, 
a negative slope can still be seen for intermediate coupling strengths, \emph{i.e.}\ $\lambda/\Omega\lesssim0.5$.

Another parameter of the model is the energy of the electronic level with respect to the Fermi level 
of the junction. Thermal stabilization, however, turns out to be rather insensitive to this parameter, 
unless additional pair-creation processes, e.g. with respect to the electrode with the lower chemical potential, 
come into play (see SI). 
The same is true, when additional electronic states are included, 
even in the presence of electron-electron interactions (cf.\ SI).

At this point, we increase the complexity of our model by coupling the vibrational mode to a heat bath. 
Fig.\ \ref{basics}d shows the average vibrational energy $\langle H_{\text{Vib}} \rangle$ of the extended 
model as a function of temperature 
for different mode-bath coupling strengths. The overall effect of the bath is to reduce the vibrational energy 
and, consequently, to narrow the range of temperatures where an increase in temperature leads to a 
reduction of the vibrational energy. 
Nevertheless, thermal stabilization is observed also in the presence of coupling to a heat bath.

So far, we considered symmetric coupling to the leads. 
Thus, the rate of transport and pair-creation processes differ only by thermal factors. Asymmetric 
molecule-lead coupling, $\Gamma_{\text{L}}\neq\Gamma_{\text{R}}$, affects the balance between transport and 
pair-creation processes, leading, for example, to vibrational rectification \cite{Hartle2010b} or 
mode-selective vibrational excitation \cite{Hartle2010,Volkovich2011b}. Fig.\ \ref{basics}e shows 
the average vibrational energy for our minimal model of a molecular junction 
as a function of temperature for different asymmetry scenarios. If the lead with the higher chemical potential 
is more strongly coupled to the molecule, cooling via pair-creation processes is more effective. The corresponding 
reduction of vibrational energy has a similar effect as coupling of the mode to a heat bath. 
In the opposite case, where the lead with the higher chemical potential is less strongly coupled to the molecule, 
cooling via pair-creation processes is less effective. Consequently, the average vibrational energy at low temperatures 
increases and the stabilization regime extends over a broader range of temperatures. 
In this regime, the quenching of heating processes is the dominant mechanism for thermal stabilization, 
in particular for $\Gamma_{\text{L}}\lesssim\Gamma_{\text{R}}/100$.

\begin{figure}
\begin{tabular}{ll}
\resizebox{0.5\textwidth}{0.19\textwidth}{
\hspace{-2.cm}\includegraphics[scale=1]{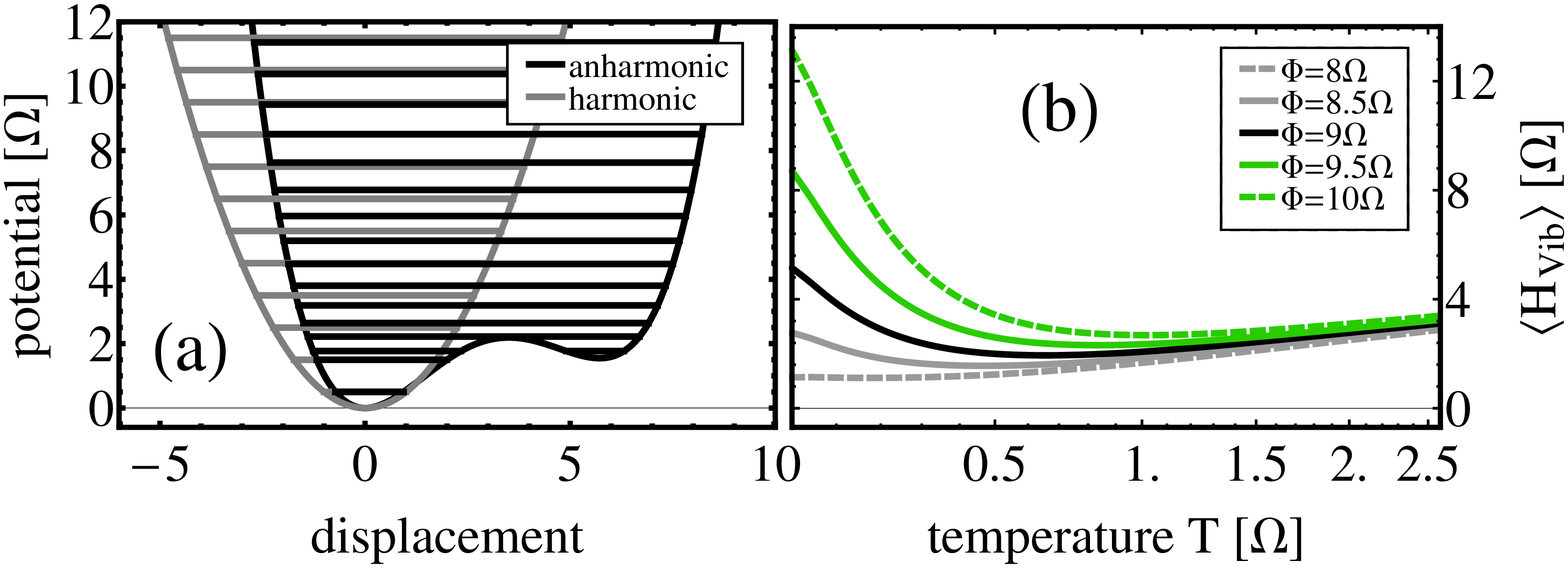}
}\\
\end{tabular}
\caption{\label{anharmonic} Panel (a): Potential of the vibrational motion for $V_2=0.136\Omega$, $V_3=0.084\Omega$ and $V_4=0.005\Omega$. 
Panel (b): Average vibrational energy $\langle H_{\text{Vib}} \rangle$ as a function of temperature for 
$\epsilon_1=3\Omega$ and the anharmonic potential shown in Panel (a). We used BM to obtain these results. 
}
\end{figure}

Next, we discuss the effect of higher-order processes and broadening due to the hybridization 
of the molecule with the leads. To this end, we employ the HQME approach and 
symmetric coupling to the leads ($\Gamma_{\text{L}}=\Gamma_{\text{R}}=\Gamma$). 
Fig.\ \ref{basics}f shows converged data for 
the average vibrational energy as a function of temperature for different molecule-lead coupling strengths. 
In the anti-adiabatic regime $\Gamma\ll\Omega$ (black line), we recover the BM results. 
Increasing the coupling strength, the average vibrational energy at low temperatures decreases. 
We attribute this behavior to broadening effects due to the hybridization of the molecule with the leads 
which have a similar effect as an increased thermal broadening. 
Accordingly, the stabilization effect is most pronounced in the anti-adiabatic regime, 
where BM theory applies.

Last but not least, we demonstrate that the phenomenon of thermal stabilization is not restricted to harmonic vibrations. 
To this end, we relax the harmonic approximation and 
consider an anharmonic potential (see black line in Fig.\ \ref{anharmonic}a). 
The corresponding average vibrational energy is depicted in Fig.\ \ref{anharmonic}b. 
Despite differences to the harmonic case, the anharmonic system still exhibits 
a pronounced regime where thermal stabilization occurs. Qualitative changes occur at low temperatures, 
because transport-induced and pair-creation processes can now take place at a variety of energies, 
redistributing the onset voltages for excitation and deexcitation processes. 
Thus, e.g., the plateau at low temperatures ($T\ll\Omega$) is less pronounced.

We conclude that molecular junctions exhibit a broad range of parameters 
where they can be stabilized by increasing the temperature of the environment. As we showed, 
the stabilization effect is robust and occurs in the weak- to intermediate-coupling regime at voltages where resonant 
pair-creation processes with a single or more vibrational quanta (cf.\ Fig.\ \ref{processes}c) 
are suppressed. This suppression is typical for low temperatures, i.e.\ $T\lesssim \Omega$. 
An increase in temperature leads to unbalanced transport-induced cooling and heating processes 
which favors the former and to a gradual reactivation of cooling by pair creation, resulting 
in a reduction of the vibrational energy. The phenomenon is more pronounced at even higher bias voltages and in 
the anti-adiabatic regime $\Gamma\lesssim\Omega$ and occurs for weak to intermediate electronic-vibrational coupling. 
We also observe it in more complex models with multiple electronic and vibrational degrees of freedom 
as well as for anharmonic vibrations. 
Our findings obtained for the specific example of molecular junctions apply similarly to other nanoelectronic 
systems, in particular ciruit QED systems where the tunneling electrons interact 
with photonic instead of vibrational degrees of freedom.

This work was supported by the German Research Foundation (DFG) and 
the German-Israeli Foundation for Scientific Research and Development (GIF).

\vspace{2cm}

\newpage
\pagebreak
\widetext
\begin{center}
\textbf{\large Supporting Information: Cooling by heating in nonequilibrium nanosystems}
\end{center}
\setcounter{equation}{0}
\setcounter{figure}{0}
\setcounter{table}{0}
\setcounter{page}{1}
\makeatletter
\renewcommand{\theequation}{S\arabic{equation}}
\renewcommand{\thefigure}{S\arabic{figure}}

\section{Influence of electronic level position}

\begin{figure}[h]
\begin{tabular}{c}
\resizebox{\newwidth}{\newheight}{
\includegraphics{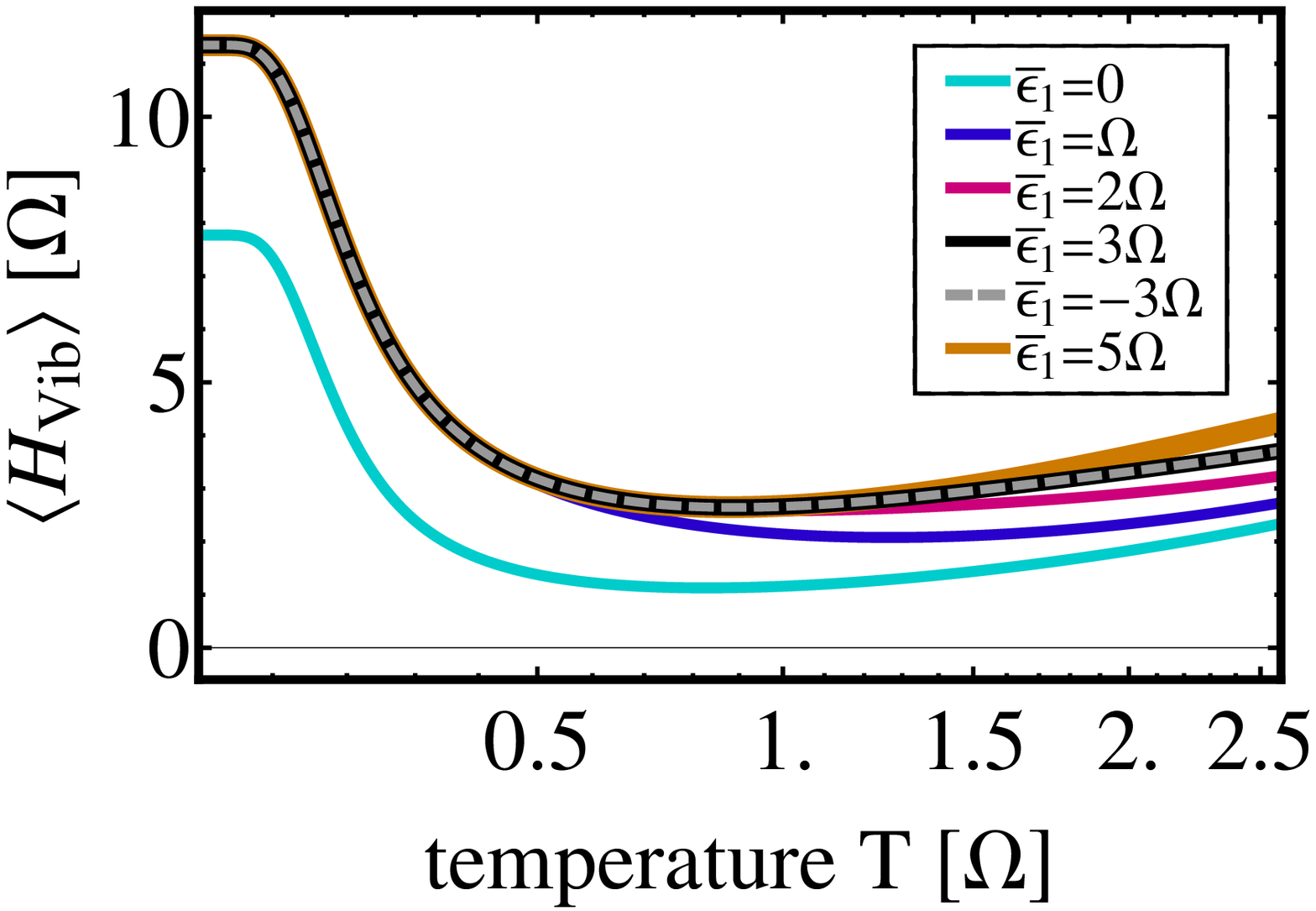}
}\\
\end{tabular}
\caption{\label{levelpos} Average vibrational energy $\langle H_{\text{Vib}} \rangle$ for our minimal model of a molecular junction 
as a function of temperature for different level positions $\overline{\epsilon}_1$. 
These results have been obtained using BM and $\Phi=2(\overline{\epsilon}_1+3\Omega/2)$. }
\end{figure}

One of the parameters of our minimal model is the position of the electronic level with respect to the Fermi level 
of the junction. Fig.\ \ref{levelpos} shows the temperature dependence of the average vibrational energy $\langle H_{\text{Vib}} \rangle$ 
for all relevant 
level positions. The effect turns out to be rather insensitive to this parameter. In fact, we obtain the same level 
of vibrational energy if we replace $\overline{\epsilon}_1$ by $-\overline{\epsilon}_1$ (black and dashed gray line). 
For higher energies (orange line), 
the results in the low and intermediate temperature regime are almost the same, including the regime where thermal stabilization occurs. 
Only if the electronic level approaches the Fermi level (turquoise line), that is if $\overline{\epsilon}_1\rightarrow0$, 
we observe a reduction of the low-temperature vibrational energy by about $50\%$ to $7.8$ at 
$\Phi=2(\overline{\epsilon}_1+3\Omega/2)$. 
Yet, the overall temperature dependence is very similar to the one for $\overline{\epsilon}_1=3\Omega$ (black line). 

\section{Influence of other electronic states}
\label{secstateSec}

Another important influence is exerted by the presence of other electronic levels. A higher-lying 
electronic level, for example, can facilitate additional inelastic processes that 
lead to a reduction of the vibrational energy at high bias voltages \cite{Hartle09,Romano10}. 
A similar behavior can be observed in the presence of electron-electron interactions (Coulomb cooling) \cite{Hartle2010b}. 
To study these effects, we add another electronic level to our model. It is distinguished from the first one only by its 
energy $\overline{\epsilon}_2=\epsilon_2-\lambda_2^2/\Omega$.

\begin{figure}[h]
\resizebox{\newwidth}{\newheight}{
\includegraphics{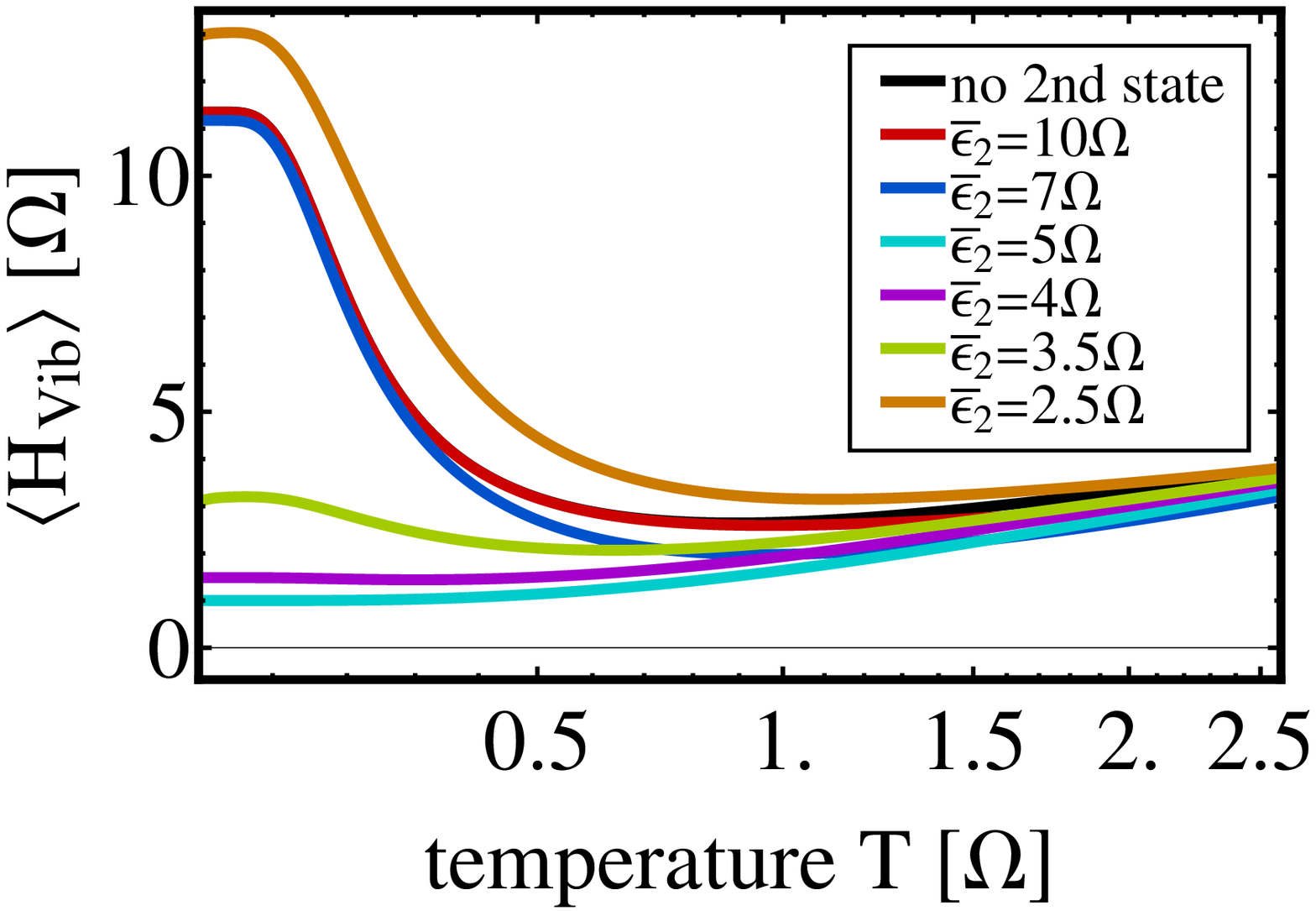}
}
\resizebox{\newwidth}{\newheight}{
\includegraphics{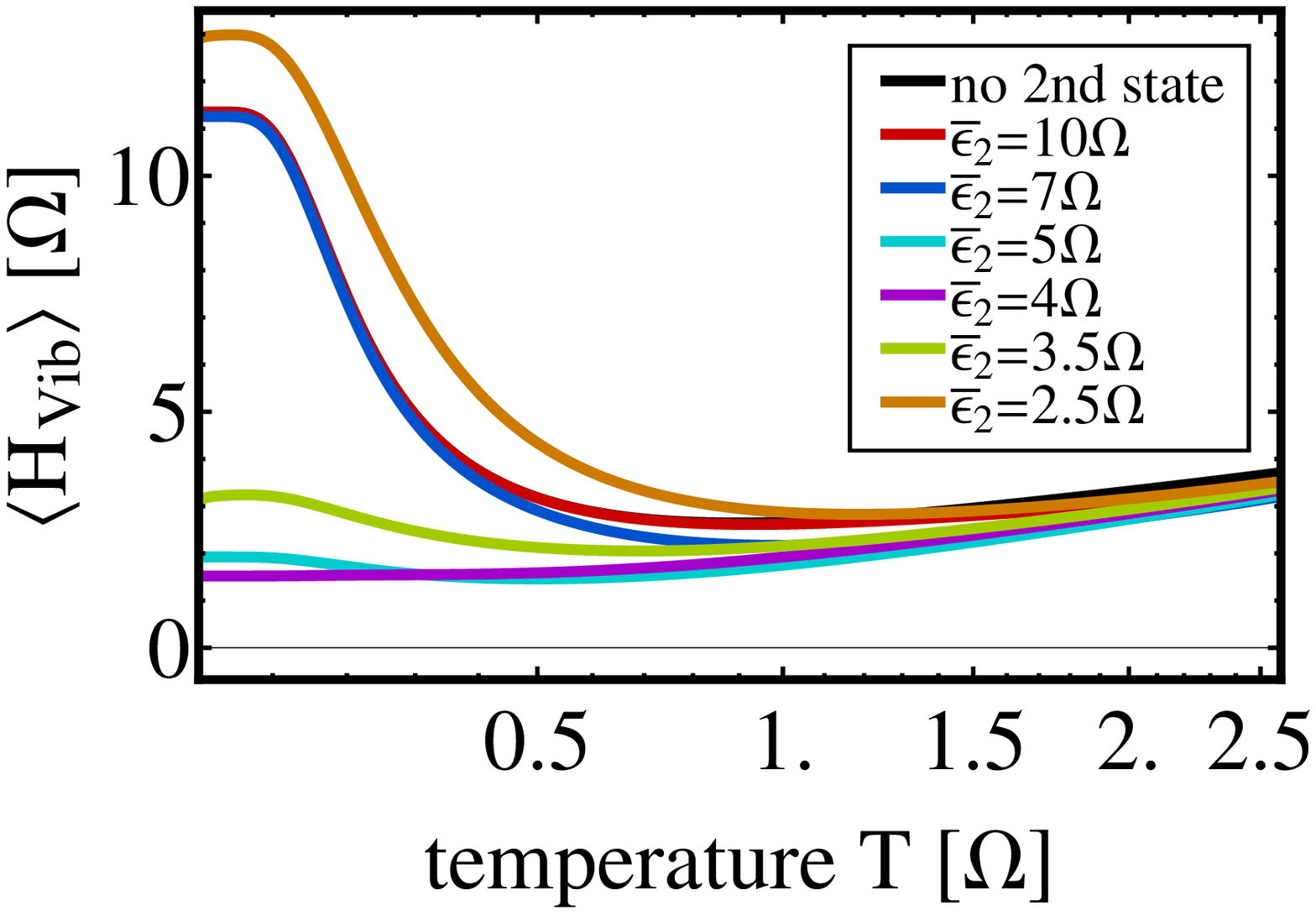}
}
\caption{\label{secondstate} Average vibrational energy $\langle H_{\text{Vib}} \rangle$ for our minimal model of a molecular junction 
 as a function of temperature, including a second electronic state 
 that is located at a set of different energies $\overline{\epsilon}_2$. 
The left/right panel shows results for $\overline{U}_{12}=0$/$\overline{U}_{12}=5\Omega$. 
These results have been obtained using BM and $\Phi=2(\overline{\epsilon}_1+3\Omega/2)=9\Omega$. }
\end{figure}

Fig.\ \ref{secondstate} shows the temperature dependence of the average vibrational energy $\langle H_{\text{Vib}} \rangle$
for different positions of the second electronic level, without electron-electron 
interactions $\overline{U}_{12}=U_{12}-2\lambda_1\lambda_2/\Omega=0$ 
(left panel of Fig.\ \ref{secondstate}) and with electron-electron interactions $\overline{U}_{12}=5\Omega$ (right panel of 
Fig.\ \ref{secondstate}). 
If the second state is too high in energy (red and blue lines), 
it has no effect. Once it approaches the bias window from above, resonant deexcitation processes 
with respect to this level become active and suppress the overall level of vibrational energy \cite{Hartle09,Romano10} 
and, consequently, the thermal stabilization effect (cf.\ turquoise line). Once it approaches the first level or becomes located 
below, the energy levels rise again and thermal stabilization is reestablished. These findings are very similar for both scenarios, 
with or without electron-electron interactions.

\end{document}